\documentclass[aps,prl,reprint]{revtex4-1}

\usepackage{graphicx,graphics,amsfonts,amsmath,amssymb}

\usepackage{color}

\begin{document}




\title{Ultracold Fermionic Feshbach Molecules of  $^{23}$Na$^{40}$K}

\author{Cheng-Hsun Wu, Jee Woo Park, Peyman Ahmadi, Sebastian Will, and Martin W. Zwierlein}
    \affiliation{MIT-Harvard Center for Ultracold Atoms, Research Laboratory of Electronics, and Department of Physics, Massachusetts Institute of Technology,
    Cambridge, Massachusetts 02139, USA }

\date{\today}

\begin{abstract}
We report on the formation of ultracold fermionic Feshbach molecules of $^{23}$Na$^{40}$K, the first fermionic molecule that is chemically stable in its ground state. The lifetime of the nearly degenerate molecular gas exceeds 100 ms in the vicinity of the Feshbach resonance. The measured dependence of the molecular binding energy on the magnetic field demonstrates the open-channel character of the molecules over a wide field range and implies significant singlet admixture. This will enable efficient transfer into the singlet vibrational ground state, resulting in a stable molecular Fermi gas with strong dipolar interactions.
\end{abstract}

\maketitle

The quest for creating and manipulating ultracold molecules has seen spectacular advances over the past decade~\cite{krem09coldmolecules, Carr:2009}. They include the production of cold molecules from laser cooled atoms~\cite{jone06}, the formation of Feshbach molecules from ultracold atomic gases~\cite{kohl06feshbachreview,chin10feshbach}, the observation of Bose-Einstein condensates of bosonic Feshbach molecules~\cite{joch03bec,grei03molbec,zwie03molBEC,ingu08varenna,kett08varenna}, and the formation of ultracold dipolar ground-state molecules~\cite{ni08polar}. Chemically stable, fermionic ground-state molecules with strong electric dipole moments would allow access to novel phases of matter in dipolar Fermi gases~\cite{Carr:2009}. So far, $^{40}$K$^{87}$Rb is the only fermionic ground state molecule that has been produced at nearly degenerate temperatures~\cite{ni08polar}. However, the gas is unstable against the exchange reaction ${\rm KRb} + {\rm KRb} \rightarrow {\rm K}_2 + {\rm Rb}_2$~\cite{ospe10chemical}. 

Here we report on the production of ultracold fermionic Feshbach molecules of $^{23}$Na$^{40}$K near a broad Feshbach resonance~\cite{Park:2012}. The NaK molecule is chemically stable in its absolute ground state~\cite{zuch10mol}. In addition, it possesses a large induced electric dipole moment of 2.72 Debye in its singlet ground state~\cite{worm81nak}, compared to 0.57 Debye for KRb~\cite{ni08polar}. Therefore, fermionic ground-state molecules of NaK can form a Fermi sea with strong, long-range anisotropic dipolar interactions, with an interaction energy greater than 30\% of the Fermi energy.
The Fesh\-bach molecules we create provide an ideal starting point for the formation of ground-state molecules. They are nearly degenerate and long-lived, with a lifetime exceeding 100 ms. We demonstrate their largely open channel character in a wide magnetic field range. The open channel has significant admixture of the electronic spin singlet state, thus opening up a direct pathway towards the absolute singlet ground state using stimulated rapid adiabatic passage (STIRAP) via a singlet excited molecular state~\cite{krem09coldmolecules, ni08polar}.

\begin{figure}[h!]
\includegraphics[width=1\columnwidth]{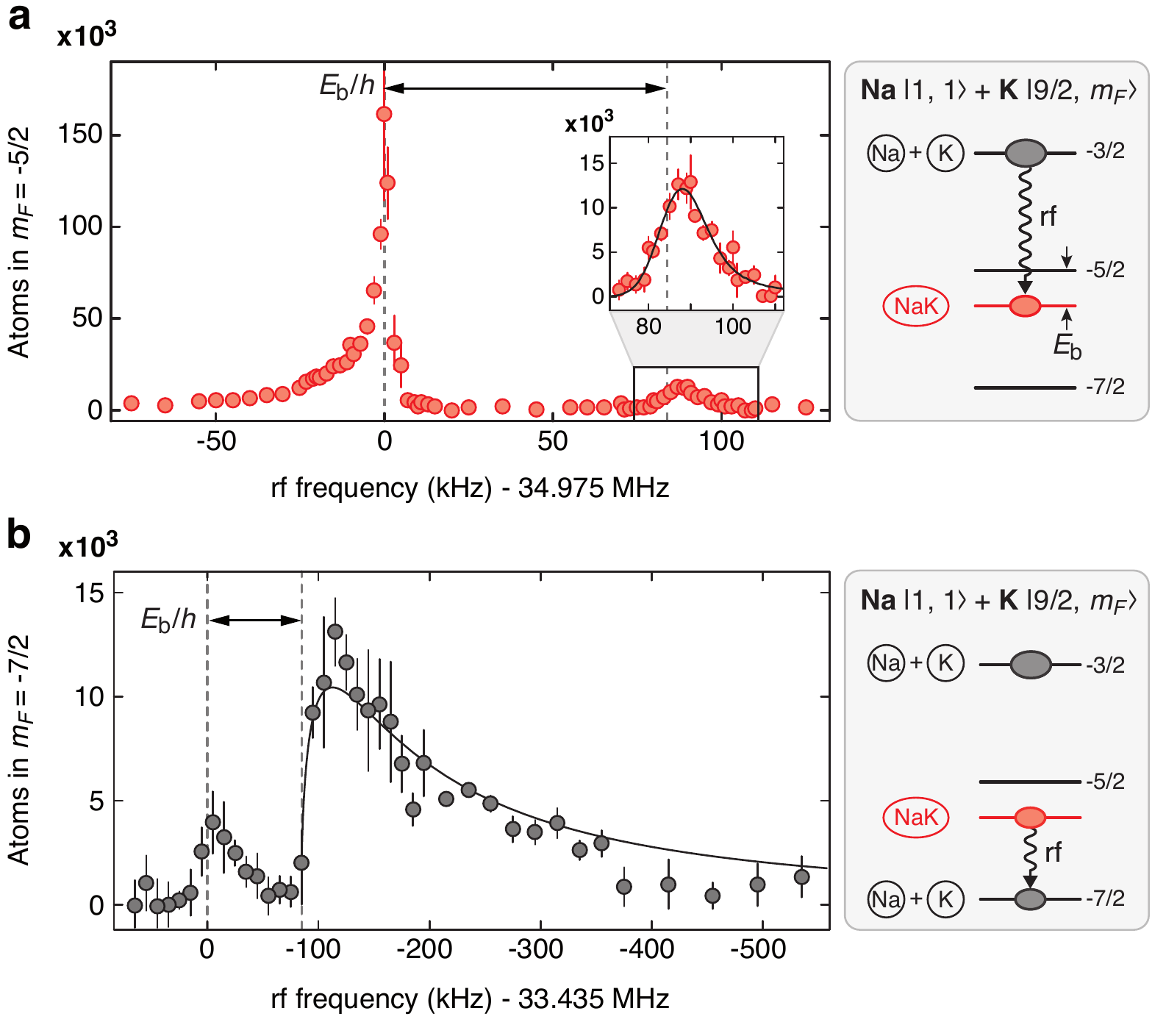}
\caption{\label{fig:1} (color online). (a) Association of fermionic Feshbach molecules. Starting with a mixture of $|1,1\rangle_{\rm Na}$ and $|9/2,-3/2\rangle_{\rm K}$ atoms, rf spectroscopy near the $|9/2,-3/2\rangle_{\rm K}$ to $|9/2,-5/2\rangle_{\rm K}$ hyperfine transition reveals free $^{40}$K atoms repulsively interacting with the $^{23}$Na bath (near zero rf offset), as well as associated molecules (near 85 kHz rf offset). A fit to the molecular association spectrum yields a binding energy of $E_{\rm b} = h \times 84(6)$ kHz. The magnetic field corresponding to the atomic transition at $34.975$ MHz was 129.4 G.
(b) Dissociation of Fesh\-bach molecules. Driving the $|9/2,-5/2\rangle_{\rm K} $ to $ |9/2,-7/2\rangle_{\rm K}$ transition yields the molecular dissociation spectrum, showing a sharp onset at an rf frequency shifted by $E_\mathrm{b}/h$ from the atomic transition. The solid line shows the fit of a model $\propto\theta(\nu - \nu_\mathrm{b})\sqrt{ \nu - \nu_\mathrm{b}}/\nu^2$~\cite{kett08varenna}, yielding $\nu_\mathrm{b} = E_\mathrm{b}/h = 85(8)$ kHz.}
\end{figure}

To form Feshbach molecules, we prepare a Bose-Fermi mixture of $^{23}$Na and $^{40}$K with $150 \times 10^3$ atoms each in a crossed optical dipole trap as has been described previously~\cite{wu11,Park:2012}\footnote{The trap frequencies for $^{40}$K [$^{23}$Na] are $\nu_x = 108(5)$ Hz [$85 (5)$ Hz], $\nu_y = 127(5)$ Hz [$ 101 (6)$ Hz] and $\nu_z = 187(4)$ Hz [$148(3)$ Hz]. The differential gravitational sag is only 4 $\mu$m}.
Sodium is in its absolute hyperfine ground state $|F,m_F\rangle = |1,1\rangle$, which forms a stable mixture with any hyperfine state of $^{40}$K in the $F = 9/2$ lower hyperfine manifold. Feshbach spectroscopy has revealed a rich pattern of interspecies Feshbach resonances~\cite{Park:2012}. One particular resonance stands out due to its exceptional width of $30\,\rm G$ at $140 \,\rm G$ between $|1,1\rangle_{\rm Na}$ and $|9/2,-5/2\rangle_{\rm K}$ atoms. Expecting open-channel dominated molecules with potentially long lifetimes, we use this resonance to form fermionic Feshbach molecules.

There are several established methods to associate Feshbach molecules~\cite{kett08varenna}. For wide Feshbach resonances, where the molecular wavefunction can have a large extent and offer good overlap with two unbound atoms, a particularly clean way to form molecules is via radiofrequency (rf) association~\cite{ospe06hetero,zirb08hetero,Zirbel:2008:2,Klempt:2008}. Our approach is schematically shown in Fig.~\ref{fig:1}(a).

Here, $^{23}$Na and $^{40}$K atoms are initially prepared in the non-resonant hyperfine states $|1,1\rangle_{\rm Na}$ and $|9/2,-3/2\rangle_{\rm K}$. For optimized phase space overlap between the two species \cite{Zirbel:2008:2}, we choose a temperature $T$ of the mixture that is close to the critical temperature $T_{\rm C}$ for Bose-Einstein condensation of sodium, corresponding to \mbox{$T/T_{\rm F} \approx 0.4$} for $^{40}$K. Via a radiofrequency pulse~\footnote{A 1.2 ms long Blackman pulse~\cite{Harris:1978} is used, allowing for Fourier-limited spectroscopy given the magnetic field stability of $\pm3$ mG that translates into about 3 kHz $\sigma$-width of transitions.}
near the $\left|9/2,-3/2\right>_{\rm K}$ to $\left|9/2,-5/2\right>_{\rm K}$ hyperfine transition, we then transfer the unbound $^{40}$K atoms into the molecular bound state with sodium.

A typical rf spectrum at a magnetic field on the molecular side of the Feshbach resonance is shown in Fig.~\ref{fig:1}(a). Two features are observed: An atomic peak near the unperturbed hyperfine transition, and a molecular peak arising from $^{40}$K atoms that have been rf associated into NaK molecules. The distance between the atomic peak and the onset of the molecular feature yields the binding energy $E_{\rm b}$. Direct absorption imaging of the large Feshbach molecules is possible using light resonant with the atomic transition~\cite{zwie03molBEC,ospe06hetero,zirb08hetero,Zirbel:2008:2,Klempt:2008}. We typically detect up to $20\times 10^3$ molecules, corresponding to a conversion efficiency of $^{40}$K atoms into molecules of about $15\,\%$.

The atomic feature at the hyperfine transition from $\left|9/2,-3/2\right>_{\rm K}$ to $\left|9/2,-5/2\right>_{\rm K}$ shows a strong mean-field shift towards lower frequencies, which is a direct signature of the repulsive interactions between $\left|1,1\right>_{\rm Na}$ and $\left|9/2,-5/2\right>_{\rm K}$ atoms on the repulsive branch of the Feshbach resonance. The free-free spectrum is directly proportional to the number of fermions that experience a given density of bosons. The long tail towards lower frequencies is caused by a small fraction of condensed bosons (about $10\%$) present in the mixture.

For accurate extraction of the molecular binding energy $E_{\rm b}$, the functional form of the molecular association spectrum is required. The lineshape can be modeled via Fermi's Golden Rule as
\begin{equation}
  \Gamma_{\rm mol}(\nu) \propto {\cal F}(h\nu - E_{\rm b})\, p(h\nu - E_{\rm b})
\end{equation}
where  ${\cal F}(\epsilon) \propto (1+\epsilon/E_{\rm b})^{-2}$ is the Franck-Condon factor between the wavefunctions of an unbound Na-K atom pair and a bound Feshbach molecule of binding energy $E_{\rm b}$~\cite{chin05rf}, and $p(\epsilon)$ is the probability density to find a pair of potassium and sodium atoms with relative kinetic energy $\epsilon$. For thermal clouds, $p(\epsilon) = \rho(\epsilon)\,\lambda_M^3 e^{-\frac{\mu}{M}\frac{\epsilon}{k_{\rm B} T}}$, where $\rho(\epsilon)$ is the density of states with relative kinetic energy $\epsilon$, $\lambda_M = \sqrt{\frac{2 \pi \hbar^2}{M k_{\rm B} T}}$ is the thermal de Broglie wavelength for an atom pair, and $M = m_{\rm Na}+m_{\rm K}$ and $\mu = \frac{m_{\rm Na}m_{\rm K}}{m_{\rm Na}+m_{\rm K}}$ are the total and reduced mass, respectively. The inset in Fig.~\ref{fig:1}(a) shows a fit of this lineshape to the molecular feature, giving a binding energy of $E_{\rm b} = h \times 84(6)$ kHz at a magnetic field of $B=129.4$ G. The width reflects the initial distribution of relative momenta between bosons and fermions and, indeed, is found to be compatible with the temperature of the sodium and potassium clouds.

A complementary demonstration of molecule formation is the observation of a dissociation spectrum~\cite{rega03mol}, as shown in Fig.~\ref{fig:1}(b). To dissociate molecules, we drive the  $\left|9/2,-5/2\right>_{\rm K}$ to $\left|9/2,-7/2\right>_{\rm K}$ transition after rf association of molecules. The onset of transfer is observed at $E_{\rm b}/h = 85(8)$ kHz, in good agreement with the association threshold in Fig.~\ref{fig:1}(a). We make use of molecule dissociation in the lifetime measurements discussed below to ensure the exclusive detection of molecules.

\begin{figure}
\includegraphics[width=1\columnwidth]{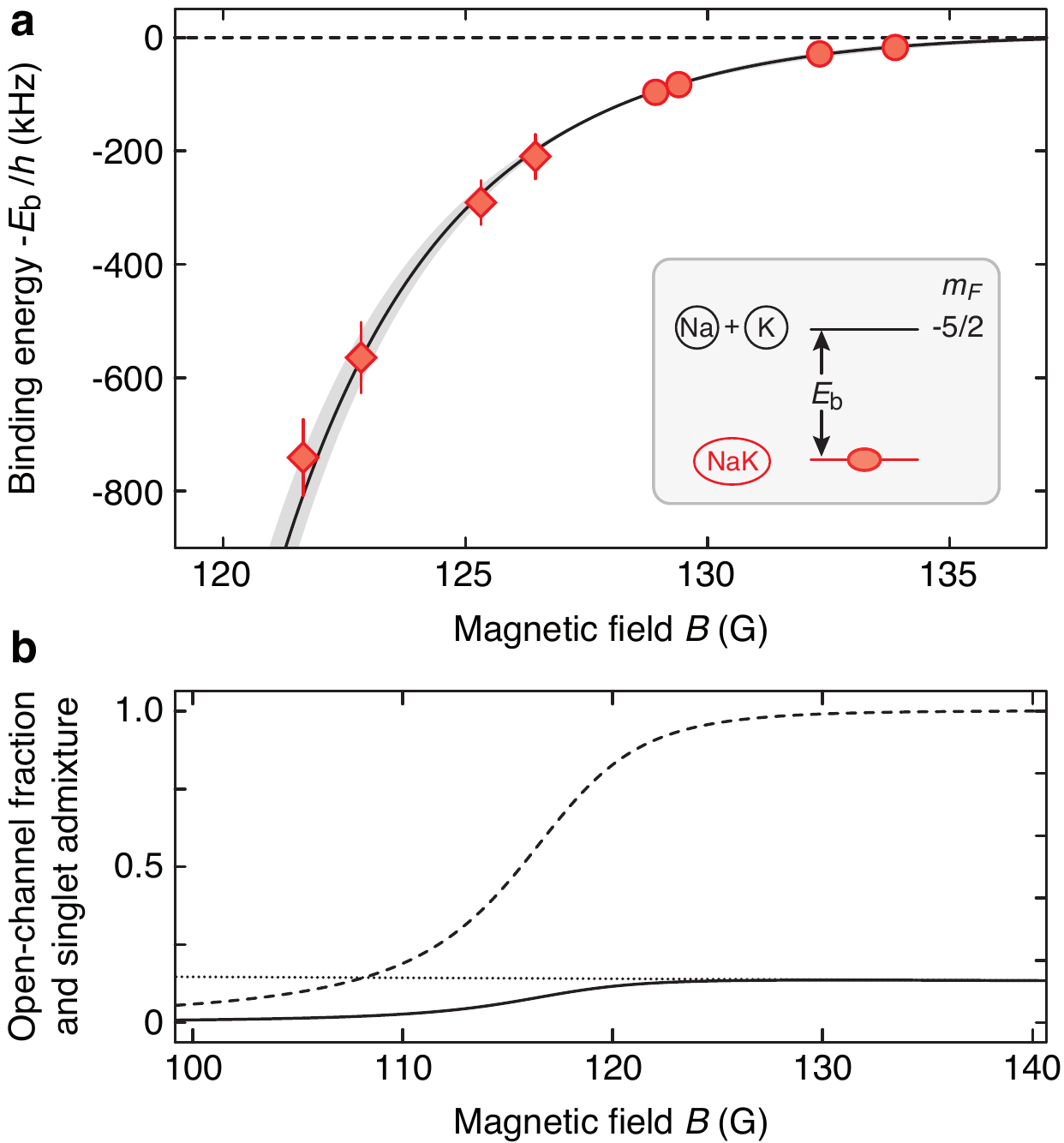}
\caption{\label{fig:2} (color online). (a) Binding energy of NaK Feshbach molecules at the wide Feshbach resonance between $|1, 1 \rangle_{\rm Na}$ and $|9/2, -5/2 \rangle_{\rm K}$. Binding energies smaller than $200$ kHz (circles) are obtained by direct detection of molecules (see Fig.~1(a)), while larger binding energies (diamonds) with weaker free-bound coupling are measured by detecting simultaneous atom loss in $|1, 1 \rangle_{\rm Na}$ and $|9/2, -3/2 \rangle_{\rm K}$. The solid line is a one-parameter fit to the model described in the text, and the shaded region displays the uncertainty. (b) Open-channel fraction (dashed line) and singlet admixture of the molecular state (solid line), derived from the binding energy curve in (a), and singlet admixture of unbound pairs (dotted line).
}
\end{figure}

We have studied the molecular binding energies as a function of magnetic field (see Fig.~\ref{fig:2}(a)). The approximately quadratic dependence of the binding energy on magnetic field reflects the open-channel character of the molecular state over a wide field range, where the binding energy follows the law $E_{\rm b} \approx \frac{\hbar^2}{2\mu(a - \bar{a})^2}$, with $\bar{a} = 51 a_0$ the mean scattering length for NaK and $a(B) \approx a_{\rm bg}\left(1 + \frac{\Delta B}{B-B_0}\right)$ the dependence of the scattering length on magnetic field near the resonance~\cite{chin10feshbach}. Since the background scattering length $a_{\rm bg} = -690^{+90}_{-130}$ is large and negative~\cite{Park:2012}\footnote{From the potential in~\cite{gerd08nak}, modified to yield the triplet binding energy found in~\cite{Park:2012}, we obtain a triplet scattering length $a_t = -805^{+100}_{-150}$, where the error bar reflects not only the uncertainty in the triplet binding energy, as in~\cite{Park:2012}, but also the uncertainty in the $C_6$ coefficient~\cite{gerd08nak}.}, the closed channel molecular state is predominantly coupled to a virtual state~\cite{marc04res,kemp04} of energy $E_{\rm vs} = \frac{\hbar^2}{2\mu(a_{\rm bg} - \bar{a})} = k_{\rm B}\times 11(3)\,\rm \mu K$.
In this case, the binding energy $E_{\rm b}$ is given by solving
\begin{equation}
  E_{\rm b} + E_{\rm c}(B) - \frac{\frac{1}{2} A_{\rm vs}}{E_{\rm vs}\left(1 + \sqrt{\frac{E_{\rm b}}{E_{\rm vs}}}\right)} = 0.
\end{equation}
Here, $E_{\rm c}(B)$ is the magnetic field dependent energy of the closed channel molecule relative to the scattering threshold, known from the asymptotic bound state model~\cite{Park:2012}, and $A_{\rm vs}$ is the squared magnitude of the coupling matrix element between the closed channel molecular state and the virtual state~\cite{marc04res,kemp04}.
With $A_{\rm vs}$ as the only fit parameter, we obtain excellent agreement with the data, finding $A_{\rm vs} = (h \times 5.2(6) \,\rm MHz)^2$. The uncertainty in the background scattering length translates into the shaded uncertainty region in Fig.~\ref{fig:2}(a). The Feshbach resonance position is found to be $B_0 = 139.7^{+2.1}_{-1.4}\,\rm G$, its width (distance from resonance to the zero-crossing of the scattering length) is $\Delta B = 29(2)\,\rm G$.

From the change of the binding energy with magnetic field we can directly deduce the closed-channel fraction $Z \approx \frac{\partial E_{\rm b}}{\partial E_{\rm c}} = \frac{1}{\Delta\mu}\frac{\partial E_{\rm b}}{\partial B}$ \cite{chin10feshbach}, where $\Delta\mu = 2.4 \mu_{\rm B}$ is the difference between the magnetic moments of the closed channel molecular state and the two free atoms near resonance. In Fig.~\ref{fig:2}(b), we show the open-channel fraction $1-Z$ along with the singlet admixture of the molecular state. Although the closed channel molecular state is a bound state in the $a(1)^3\Sigma^+$ triplet potential, the strong coupling to the entrance channel, which is $14\%$ singlet, gives the Feshbach molecules a mixed singlet-triplet character. This will allow for direct two-photon coupling to the $v=0$, $X(1)^1\Sigma^+$ singlet vibrational ground state via the singlet excited $B(1)^1\Pi$ state.

\begin{figure}
\includegraphics[width=1\columnwidth]{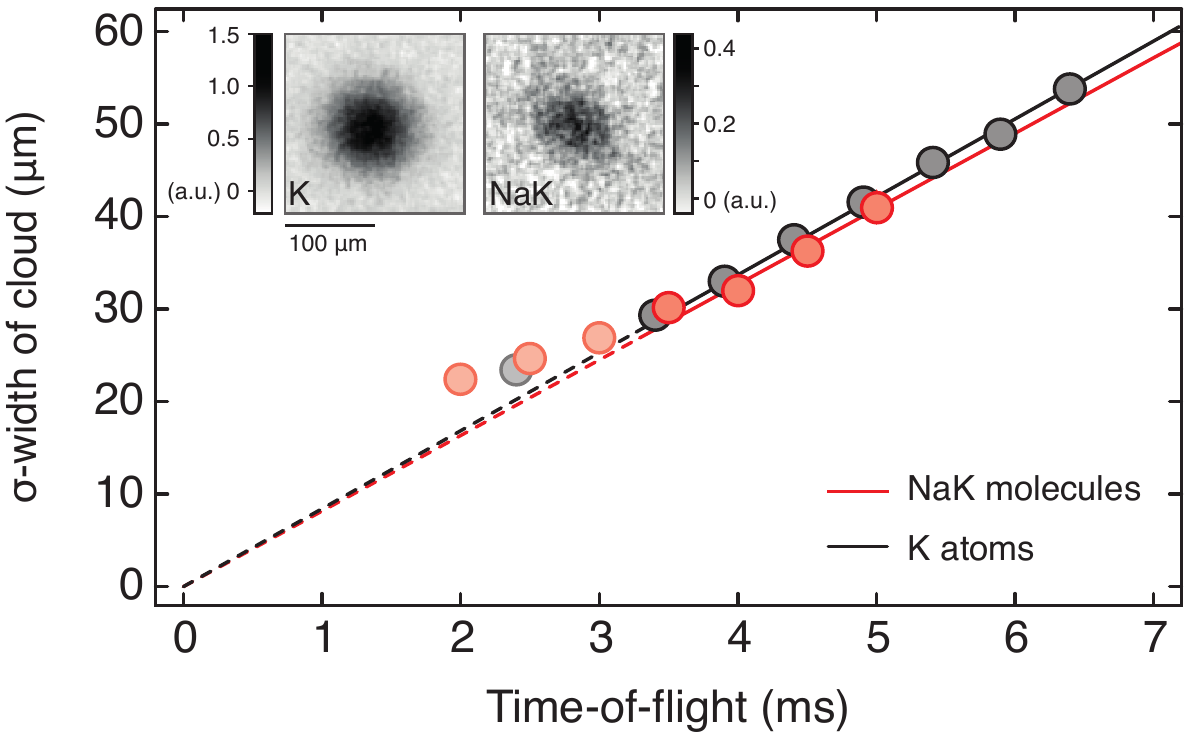}
\caption{\label{fig:3} (color online). Time-of-flight expansion of leftover $^{40}$K atoms and associated molecules at $129.4$ G (binding energy $h \times 84\,\rm kHz$) right after rf association. Assuming a classical gas model, the width of the cloud corresponds to an average kinetic energy of $340\,\rm nK$ for the $^{40}$K cloud and $500 \,\rm nK$ for the molecular cloud. Free atoms are transferred and detected in the $m_F = -1/2$ state, while potassium atoms bound in NaK molecules are imaged in the $m_F=-5/2$ state. The insets show  an average of 20 absorption images for K atoms (NaK molecules) after $4.5$ ms ($4.0$ ms) time-of-flight.
}
\end{figure}

To obtain an estimate for the temperature of the molecules we study their time-of-flight expansion right after association at $E_{\rm b} = h \times 84$ kHz. We measure an effective temperature (average kinetic energy) of 500 nK (see Fig.~\ref{fig:3}) for the molecules. From their number and trapping frequencies, this corresponds to a degeneracy factor of $T/T_{\rm F, mol} \approx 1.7$. The molecular gas is more energetic than the sodium and potassium bath, which has a temperature of $220$ nK. This partially reflects the broader momentum distribution of the $^{40}$K Fermi gas at $T/T_{\rm F}\approx 0.4$ due to Pauli pressure. Indeed, the momentum distribution of molecules after rf association is a convolution of the momentum distributions of the bosonic and fermionic clouds. At zero temperature, with bosons condensed and fermions fully occupying the Fermi sea, molecules would inherit the broad momentum distribution of the atomic Fermi gas. We also observe heating of the molecular cloud in the presence of sodium, likely due to three-body losses, that can occur already during the association pulse.

\begin{figure}
\includegraphics[width=1\columnwidth]{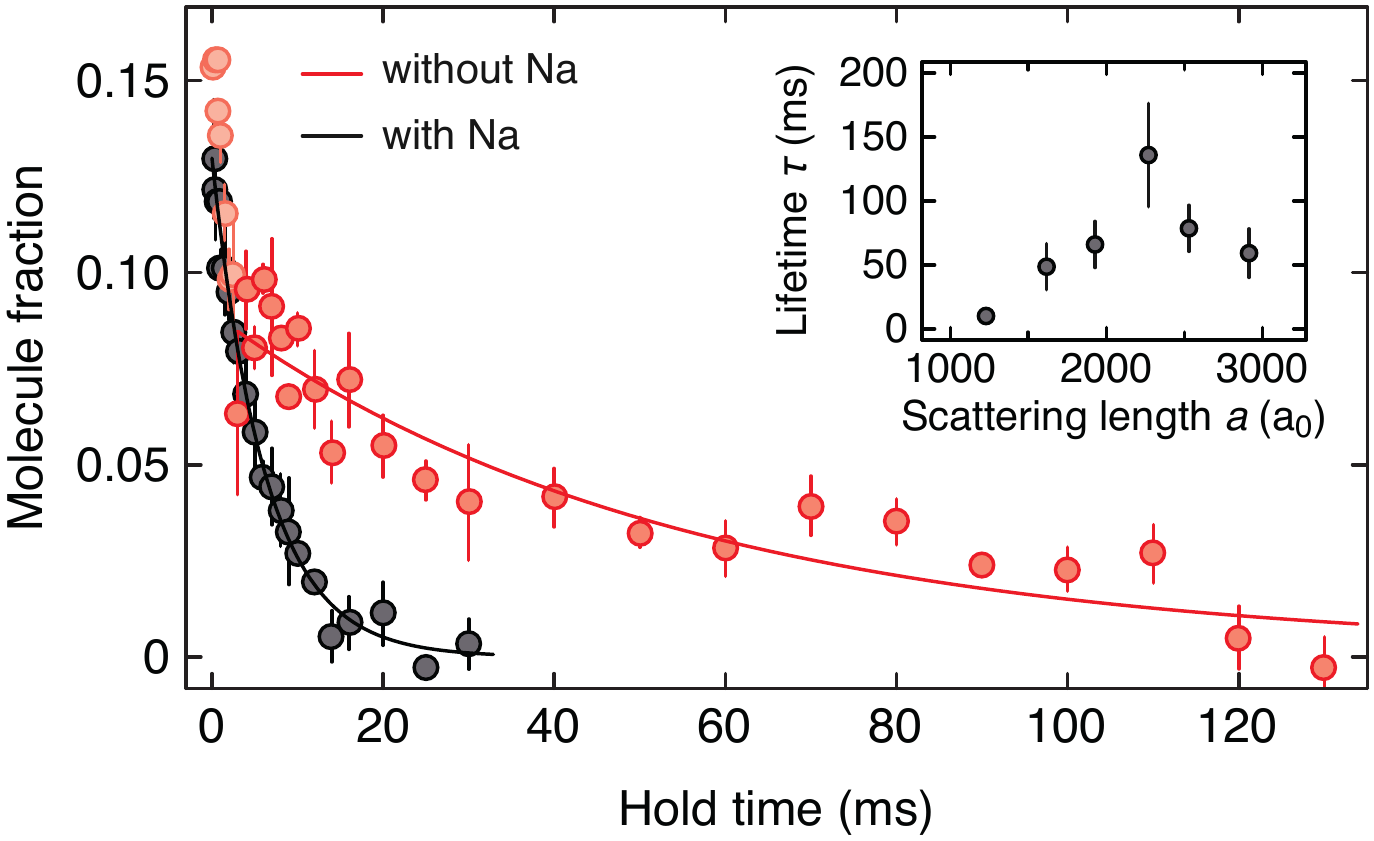}
\caption{\label{fig:4} (color online). Lifetime of fermionic Feshbach molecules. The measurement is performed at a magnetic field of $132.2$ G, corresponding to a binding energy of $32$ kHz. The red (black) data points show the number of molecules as a function of hold time with (without) removing sodium from the dipole trap directly after association. Solid lines correspond to exponential fits yielding a $1/e$-time of $54(13)\, \rm ms$ and $8(1)\, \rm ms$, respectively. The inset shows the lifetime as a function of the scattering length between bosons and fermions. The maximum observed lifetime is $140(40) \,\rm ms$.}
\end{figure}

Fig.~\ref{fig:4} shows the lifetime of molecules close to the Fesh\-bach resonance at a binding energy of $h \times 32\,\rm kHz$. With the bosonic species still present, the lifetime is about 8 ms. Considering that the bosons can resonantly scatter with the fermions bound in the NaK molecules, this is already an impressive lifetime. However, the bosons can be selectively removed from the optical trap as its depth is species selective, being more confining for NaK molecules and $^{40}$K atoms than for $^{23}$Na. With bosons removed, the lifetime increases significantly to 54 ms, and further up to 140(40) ms at a binding energy of $h\times 25\,\rm kHz$ (see inset in Fig.~\ref{fig:4}). The lifetime is likely limited by the presence of leftover $^{40}$K atoms in the $|9/2,-3/2\rangle_{\rm K}$ state. Their average density is $n_{\rm A} = \frac{1}{N} \int n^2\,{\rm d}^3 r = 1\times 10^{12} \,\rm cm^{-3}$, yielding a loss rate of $\beta = 8(2)\times 10^{-12} \,\rm cm^3/s$. This loss rate is an order of magnitude lower than the corresponding rate found for collisions of KRb Feshbach molecules with distinguishable atoms, and three times lower than the smallest rate measured for KRb Feshbach molecules with any collision partner~\cite{zirb08hetero}. The loss rate even rivals that found for bosonic Feshbach molecules formed in a heteronuclear Fermi-Fermi mixture of LiK~\cite{voig09fermi}.

For collisions with distinguishable atoms, the molecular loss rate is expected to decrease like $\beta \propto 1/a$ due to the reduced wavefunction overlap with lower-lying vibrational states~\cite{dincao2008decay}. While we observe a dramatic increase in the lifetime with scattering length for $a\lesssim 2300\, a_0$ (binding energies $E_{\rm b}\gtrsim h \times 25 \,\rm kHz$), the lifetime decreases again for even larger scattering lengths, possibly due to thermal dissociation when the binding energy becomes comparable to the temperature.

In conclusion, we have created ultracold fermionic Fesh\-bach molecules of NaK, which are chemically stable in their absolute ground state. The molecular gas is formed close to degeneracy and found to be long-lived near the Feshbach resonance. As revealed by the binding energy measurements, the Feshbach molecules are open-channel dominated over a wide magnetic field range, where they possess significant singlet character. These are ideal starting conditions for a two-photon transfer into the absolute singlet vibrational ground state~\cite{ni08polar}. Combined with the large induced dipole moment of $2.72$ Debye, NaK presents us with a unique opportunity to create a stable, strongly dipolar Fermi gas with dominating long-range anisotropic interactions.

We would like to thank Tobias Tiecke for fruitful discussions, and Ibon Santiago for experimental support in the early stages of the experiment. This work was supported by the NSF, AFOSR-MURI, AFOSR-PECASE, a grant from the Army Research Office with funding from the DARPA OLE program and the David and Lucille Packard Foundation.

\bibliographystyle{apsrev4-1}

\end{document}